\documentclass[aps,prd,twocolumn,a4paper]{revtex4}
\usepackage{ulem}
\usepackage{color}
\usepackage{graphicx}
\usepackage{epsfig}
\def\nn {\nonumber}

\newcommand{\be}{\begin{equation}}
\newcommand{\ee}{\end{equation}}
\newcommand{\bea}{\begin{eqnarray}}
\newcommand{\eea}{\end{eqnarray}}

\newcommand{\om}{\omega}

\newcommand{\ov}{\overline}

\newcommand{\ls}{l \!\!\! /}

\newcommand{\vk}{\vec k}
\newcommand{\vl}{\vec l}

\newcommand{\F}{F_\pi}

\begin{document}

\title{Shear viscosity due to the Landau damping from quark-pion interaction}
\author{Sabyasachi Ghosh$^{1,2}$, Anirban Lahiri$^2$, Sarbani Majumder$^2$, Rajarshi Ray$^2$, Sanjay K. Ghosh$^2$}
\affiliation{$^1$Instituto de Fisica Teorica, Universidade Estadual Paulista, 
Rua Dr. Bento Teobaldo Ferraz, 271, 01140-070 Sao Paulo, SP, Brazil}
\affiliation{$^2$Center for Astroparticle Physics and Space Science, 
Bose Institute, Block EN, Sector V, Salt Lake, Kolkata 700091, India}


\begin{abstract}
We have calculated the shear viscosity coefficient $\eta$ of the strongly 
interacting matter in the relaxation time approximation, 
where a quasi particle description of quarks with its dynamical mass
is considered from NJL model.
Due to the thermodynamic scattering of quarks
with pseudo scalar type condensate (i.e. pion), a non zero 
Landau damping will be acquired by the propagating quarks.
This Landau damping may be obtained from the Landau cut
contribution of the in-medium self-energy of quark-pion loop,
which is evaluated in the framework of real-time thermal field theory.
\end{abstract}

\maketitle
From the basic idea of the QCD asymptotic freedom at
high temperatures and densities, a weakly interacting
quark gluon plasma (QGP) is naturally expected to be produced
in the experiments of heavy ion collision (HIC). 
However, the experimental data from RHIC, especially the measured
elliptic flow
indicates that nuclear matter as a strongly interacting 
liquid instead of a weakly interacting gas.
The recent hydrodynamical calculations~\cite{Romatschke,Heinz}
as well as some calculations of kinetic transport theory~\cite{Xu, Greco} 
conclude that the matter, 
produced in HIC, must have very small shear viscosity.
The shear viscosity of the fluid is generally quantified by the
the coefficient $\eta$ and it
physically interprets the ability to transfer momentum over a
distance of mean free path. Hence the lower values of $\eta$
means the constituents of the matter interact strongly to 
transfer the momentum easily.
Whereas a weakly
interacting system must have large $\eta$ because 
in this case the momentum transfer between
the constituents become strenuous.

Several theoretical attempts~\cite{Arnold,Nakamura,Meyer,Prakash,Dobado,
Nakano,Itakura,Nicola,Muronga,Weise,Gavin,SSS,Plumari,Krein,Csernai,Purnendu,
Gyulassy,Hufner,Redlich_NPA,Bass,Cassing} 
are taken to calculate the $\eta$ of the strongly interacting matter at very
high~\cite{Arnold}, intermediate~\cite{Nakamura,Meyer} and
low temperature~\cite{Prakash,Dobado,Nakano,Itakura,Nicola,Muronga,Weise,Gavin,SSS}, 
where some special attentions are drawn on the smallness of 
its original value with respect to its
lower bound ($\eta=\frac{s}{4\pi}$, where $s$ is entropy density),
commonly known to as the KSS bound~\cite{KSS}.
The most interesting fact, which has been added with the recent theoretical 
understanding of $\eta$ for strongly interacting matter, is that the $\eta/s$
may reach a minimum in the vicinity of a phase transition~\cite{Csernai,Purnendu,
Gyulassy,Hufner,Redlich_NPA} (see also \cite{Chen_Tc}) like the liquid-gas
phase transition of certain materials e.g. Nitrogen, Helium or Water.
These investigations demand a better understanding to zoom in on 
the temperature ($T$) dependence of $\eta$ 
of the strongly interacting matter near the phase transition. 
Inspiring by this motivation, in this brief report we have addressed the $\eta (T)$
due to forward and backward scattering of quark-pion interaction.

In the relaxation time approximation, the $\eta$ of the quark~\cite{Redlich_NPA} 
and pion~\cite{Gavin,SSS} medium (for $\mu=0$) can be expressed as
\bea
\eta&=&\frac{8\beta}{5}\int \frac{d^3\vk}{(2\pi)^3}\frac{\vk^4}{\om_Q^2}
\frac{n_Q(1-n_Q)}{\Gamma_Q}
\nn\\
&&+\frac{\beta}{5}\int \frac{d^3\vk}{(2\pi)^3}\frac{\vk^4}{\om_\pi^2}
\frac{n_\pi(1+n_\pi)}{\Gamma_\pi}
\label{eta_final}
\eea
where $n_{Q}=\frac{1}{e^{\beta\om_{Q}}+1}$ and 
$n_{\pi}=\frac{1}{e^{\beta\om_{\pi}}-1}$ are respectively Fermi-Dirac distribution
of quark and Bose-Einstein distribution of pion 
with $\om_Q=\sqrt{\vk^2+M_Q^2}$ and $\om_\pi=\sqrt{\vk^2+m_\pi^2}$. 
The $\Gamma_{Q}$ and $\Gamma_\pi$ are Landau damping of quark and pion respectively.
Following the quasi particle description of Nambu-Jona-Lasinio 
(NJL) model~\cite{NJL_rev}, the dynamical quark mass $M_Q$
is considered and it is generated due to quark condensate
\be 
\langle{\ov \psi}_f\psi_f\rangle=-\frac{M_Q-m_Q}{2G}
\ee
where $m_Q$ is the current quark mass.
In the medium, above relation become (for $\mu=0$)
\be
M_Q=m_Q+4N_fN_cG\int\frac{d^3\vk}{(2\pi)^3}\frac{M_Q}{\om_Q}(1-2n_Q)~.
\label{gap}
\ee
This relation shows that the constituent quark mass tends to be
the current quark mass at very high temperature
where the non-zero quark condensate becomes small.
\begin{figure}
\includegraphics[scale=0.5]{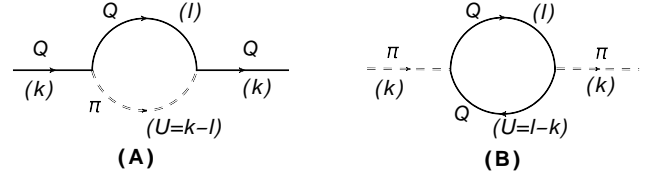}
\caption{The diagram of quark {\bf (A)} and
pion {\bf (B)} self-energy for quark-pion and quark-anti quark
loops respectively.}
\label{gm_diagram}
\end{figure}

This Landau damping $\Gamma_Q$ and $\Gamma_\pi$ may be estimated from the 
self-energy graphs of quark and pion at finite temperature for quark-pion
and quark-anti quark loops respectively.
These are respectively expressed as
\be
\Gamma_Q=-{\rm Im}{\Sigma}^R(k_0=\sqrt{\vk^2+M_Q^2},\vk)
\ee
and
\be
\Gamma_\pi=-\frac{1}{m_\pi}{\rm Im}{\Pi}^R(k_0=\sqrt{\vk^2+m_\pi^2},\vk)
\ee
where ${\Sigma}^R$ and $\Pi^R$ are respectively retarded part of 
quark and pion self-energy at finite temperature. Their diagrammatic
representations are shown in Fig.\ref{gm_diagram}{\bf (A)} and 
{\bf (B)} respectively.
Following the real-time formalism of thermal field theory, the
retarded part of in-medium quark self energy for quark-pion loop
is given by~\cite{S_thesis}
\bea
&&{\Sigma}^R(k_0,\vk)=\int\frac{d^3{\vec l}}{(2\pi)^3}\frac{1}{4\om^l_Q\om^U_\pi}
[\frac{(1-n^l_Q)L^Q_1+n^U_\pi L^Q_3}{k_0 -\om^l_Q-\om^U_\pi+i\eta}
\nn\\
&&+\frac{n^l_Q L^Q_1+n^U_\pi L^Q_4}{k_0-\om^l_Q+\om^U_\pi+i\eta} 
+ \frac{-n^l_Q L^Q_2 -n^U_\pi L^Q_3}{k_0 +\om^l_Q-\om^U_\pi+i\eta}
\nn\\ 
&&~~~~~~~+\frac{n^l_QL^Q_2 +(-1-n^U_\pi)L^Q_4}{k_0 +\om^l_Q+\om^U_\pi+i\eta}].
\label{self_pi_Q}
\eea
where $L^Q_i,i=1,..4$ denote the values of $L^Q(l_0,{\vec l})$ for
$l_0=\om^l_Q,-\om^l_Q,k_0-\om^U_\pi,k_0+\om^U_\pi$ respectively
with $\om^l_Q=\sqrt{\vk^2+M_Q^2}$ and
$\om^U_\pi=\sqrt{(\vk-\vl)^2+m_\pi^2}$. 
Here $n^l_Q(\om^l_Q)$ is Fermi-Dirac distribution function of quark
whereas $n^U_\pi(\om^U_\pi)$ denotes Bose-Einstein distribution
function of $\pi$ meson.

During extracting the imaginary part of $\Sigma^R(k_0,\vk)$ we will get four
delta functions associated with the four individual terms of Eq.~(\ref{self_pi_Q}),
which generate four different region in $k_0$-axis where the ${\rm Im}\Sigma^R(k_0,\vk)$
will be non-zero.
From the non-zero values of ${\rm Im}\Sigma^R(k_0,\vk)$ the region of 
discontinuities or branch cuts of $\Sigma^R(k_0,\vk)$ can be identified.
The regions coming from the 1st and 4th terms of (\ref{self_pi_Q}) 
are respectively $(k_0=-\infty$ to $-\sqrt{\vk^2+(m_\pi+M_Q)^2}$) and
($k_0=\sqrt{\vk^2+(m_\pi+M_Q)^2}$ to $\infty$). These are known as
unitary cuts and different kind of forward and inverse decay processes
are associated with these cut contributions~\cite{Weldon,S_thesis}. 
Similarly the regions
($k_0=-\sqrt{\vk^2+(m_\pi-M_Q)^2}$ to $0$) and $(k_0=0$ to $\sqrt{\vk^2+(m_\pi-M_Q)^2}$)
are coming from 2nd and 3rd terms respectively. These purely medium
dependent cuts are known as Landau cuts and different kind of forward and 
inverse scattering processes are physically interpreted by these cut contributions
~\cite{Weldon,S_thesis}. So the 3rd term of ${\rm Im}\Sigma^R(k_0,\vk)$ at
the on-shell mass $(k_0=\sqrt{\vk^2+M_Q^2},\vk)$ of quark is responsible
for the Landau damping $\Gamma_Q$ and it is given by~\cite{S_thesis}
\bea
\Gamma_Q&=&-{\rm Im}\Sigma^R(k_0=\sqrt{\vk^2+M_Q^2},\vk)=
[\int\frac{d^3{\vec l}}{(2\pi)^3}\frac{L^Q_2}{4\om^l_Q\om^U_\pi}
\nn\\
&&(n^l_Q+n^U_\pi)\delta(k_0+\om^l_Q-\om^U_\pi)]_{k_0=\sqrt{\vk^2+M_Q^2}}~.
\label{gm_Q}
\eea
Rearranging the statistical weight factor by
\be
(n^l_Q+n^U_\pi)=n^l_Q(1+n^U_\pi)+n^U_\pi(1-n^l_Q)~,
\label{n_l_n_U}
\ee
we can find thermalized $\pi$ and ${\ov u}$ with
Bose enhanced probability $(1+n^U_\pi)$ and Pauli blocked probability
$(1-n^l_Q)$ respectively. With the help of Eq.~(\ref{n_l_n_U}),
the physical significance of the Landau cut
contribution may expressed as follows. During the propagation of $u$ quark,
it may absorb the thermalized ${\ov u}$ from the heat bath and create a
thermalized $\pi$ in the bath (indicated by the second part of Eq.~(\ref{n_l_n_U})).
Again the thermalized $\pi$ may be absorbed by the medium and create
the thermalized ${\ov u}$ along with a propagating $u$, which is slightly
off-equilibrium with the medium (indicated by the first part of Eq.~(\ref{n_l_n_U})).

To calculate $L^Q_i$ from quark-pion interaction, let us start with free Lagrangian of quarks
and demanding the invariance properties of Lagrangian under chiral transformation,
\be 
\psi_f'={\rm exp}(\frac{i{\vec \pi}\cdot {\vec \tau}\gamma^5}{2\F})\psi_f
\ee
where the chiral angle is associated with the pion field ${\vec \pi}$ and
$\F$ is pion decay constant. Expanding up to first order of pion field, we
obtain the quark-pion interaction term~\cite{Miller, Miller_18},
\bea 
{\cal L}_{\pi QQ}&=&\frac{-iM_Q}{\F}{\ov \psi}_f{\vec \pi}\cdot {\vec \tau}\gamma^5\psi_f
\nn\\
&=&\frac{-iM_Q\gamma^5}{\F}\left(\begin{array}{c}{\ov u}~~ {\ov d} \end{array}\right)
\left(\begin{array}{c}\pi^0~~ \sqrt{2}\pi^+\\\sqrt{2}\pi^-~~ \pi^0\end{array}\right)
\left(\begin{array}{c}u \\ d \end{array}\right)~.
\eea
As we are interested to
calculate one-loop self-energy ($\Sigma^R$) 
of any quark flavor,
$u$ (say) hence we have to consider two possible loops
- $u\pi^0$ and $d\pi^+$. Due to isospin symmetry consideration in Lagrangian,
we can evaluate anyone of the loops, say $u\pi^0$ loop and then we have to 
multiply it by a isospin factor 
\be
I_F=(1)^2+(\sqrt{2})^2=3~.
\ee 
From the interaction part,
\be 
{\cal L}_{\pi^0uu}=-ig_{\pi QQ}{\ov u}\gamma_5\pi^0u,~~~
{\rm with}~g_{\pi QQ}=\frac{M_Q}{\F}
\label{Lag_piQQ}
\ee 
we can calculate $L^Q_{ab}(l_0,{\vec l})=-I_F~ g^2_{\pi QQ}(\ls -m_l)_{ab}$,
where $a, b$ are Dirac indices. For simplification we have taken the
scalar part only i.e. $L^Q(l_0,{\vec l})=I_F~ g^2_{\pi QQ}m_l$.
We have taken the parameters $m_Q=0.0056$ GeV, $M_Q=0.4$ GeV (for $T=0$), 
three momentum cut-off $\Lambda=0.588$ GeV and
corresponding $T_c=0.222$ GeV for $\mu=0$~\cite{NJL_rev}.
\begin{figure}
\includegraphics[scale=0.3]{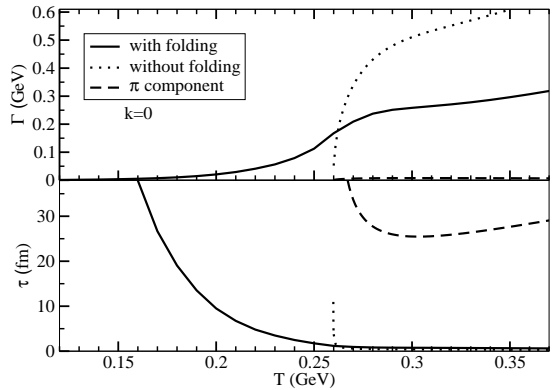}
\caption{Upper panel : The $T$ dependency of $\Gamma_Q$
with (solid line) and without (dotted line) folding by $A_\pi$ 
and $\Gamma_\pi$ (dashed line). Lower panel : The variation of corresponding 
collision time $\tau$ with temperature.}
\label{gm_T}
\end{figure}
\begin{figure}
\includegraphics[scale=0.3]{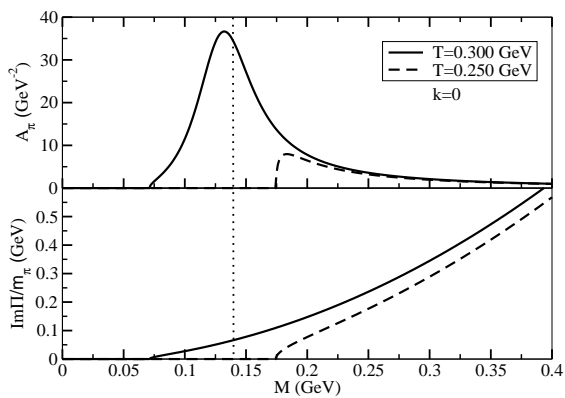}
\caption{Lower panel shows $M$ dependency of the imaginary part of pion 
self-energy for $Q{\bar Q}$ loop, which is normalized after dividing by $m_\pi$. 
Upper panel shows invariant mass distribution of pion spectral function due to
its $Q{\bar Q}$ width. Dotted line indicates the position of pion pole.}
\label{pi_spec}
\end{figure}
\begin{figure}
\includegraphics[scale=0.3]{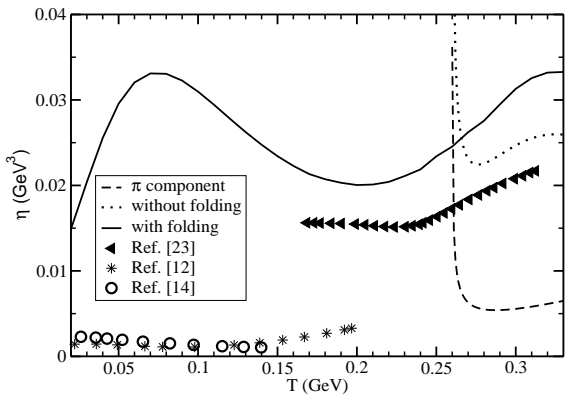}
\caption{Temperature dependence of $\eta$ due to
$\Gamma_\pi$ (dashed line), $\Gamma_Q$ without (dotted line)
and with (solid line) folding are separately shown.
The results of Ref.~\cite{Redlich_NPA}(triangles)
are attached to compare with our results (solid line).
[also the results of hadronic domain by Ref.~\cite{Nicola}(stars), 
Ref.~\cite{Weise} (open circles)].}
\label{eta_T_comp}
\end{figure}

Similar to Eq.~(\ref{self_pi_Q}), the pion self-energy $\Pi^R$ for
quark-anti quark loop is also received similar kind of form only the 
quantities $n^U_\pi$, $\om^U_\pi=\sqrt{(\vk-\vl)^2+m_\pi^2}$, 
$U=k-l$ and $L^Q_i$'s are changed to $-n^U_Q$, $\om^U_Q=\sqrt{(\vl-\vk)^2+m_Q^2}$, 
$U=-k+l$ and $L^{\pi}_i$'s respectively~\cite{S_rho,S_omega}. 
As pion on-shell mass point ($k_0=\sqrt{\vk^2+m_\pi^2},\vk$) will be
inside the unitary cut region ($k_0=\sqrt{\vk^2+(M_Q+M_Q)^2}$ to $\infty$)
of $\Pi^R$, therefore
\bea
\Gamma_\pi&=&\frac{-{\rm Im}\Pi^R(k_0=\sqrt{\vk^2+m_\pi^2},\vk)}{m_\pi}=
\frac{-1}{m_\pi}[\int\frac{d^3{\vec l}}{(2\pi)^3}\frac{L^{\pi}_1}{4\om^l_Q\om^U_Q}
\nn\\
&&(1-n^l_Q-n^U_Q)\delta(k_0+\om^l_Q-\om^U_Q)]_{k_0=\sqrt{\vk^2+m_\pi^2}}
\label{gm_pi}
\eea
where $L^{\pi}=4I_Fg^2_{\pi QQ}[M_Q^2-l^2-k\cdot l]$ can be obtained
from (\ref{Lag_piQQ}).

In Fig.~(\ref{gm_T}), we can see the temperature dependency of 
Landau damping $\Gamma$ (upper panel) and collision 
time $\tau=\frac{1}{\Gamma}$ (lower panel)
of quark (dotted line) and pion (dashed line) for their momentum $\vk=0$.
Owing to the on-shell condition, the $\Gamma_Q$ and $\Gamma_\pi$ are
received the non-zero values only in the temperature range where
$m_\pi>2M_Q$, which are clearly seen from dotted and dashed lines
respectively. A corresponding non-divergent collisional times are also 
achieved by them in the same temperature domain.
Due to decay width of $\pi\rightarrow Q{\bar Q}$, it will more
realistic to consider the pion resonance of finite width in Eq.~(\ref{gm_Q}).
The pion spectral function due to $Q{\bar Q}$ width may be defined as
\be
A_\pi(M)=\frac{1}{\pi}{\rm Im}
\left[\frac{1}{M^2-m_\pi^2+i{\rm Im}\Pi^R_{\rm vac}(k_0,\vk)}\right]
\label{A_pi}
\ee
where ${\rm Im}\Pi^R_{\rm vac}(k_0,\vk)$ is vacuum part of 
${\rm Im}\Pi^R(k_0,\vk)$ and $M=\sqrt{k_0^2-\vk^2}$.
The variation of Im$\Pi^R(M)/m_\pi$ and $A_\pi$ with $M$ for two different
temperatures are respectively shown in lower and upper panel of Fig.~(\ref{pi_spec}).
Replacing $m_\pi$ of $\Gamma_Q$ in (\ref{gm_Q}) by $M$ and then
convoluting or folding it by $A_\pi(M)$, we have~\cite{S_rho,S_omega}
\be
\Gamma_Q(m_\pi)=\frac{1}{N_\pi}\int\Gamma_Q(M)A_\pi(M)dM^2
\label{gm_Q_fold}
\ee
where $N_\pi=\int A_\pi(M)dM^2$. One should notice that
in the narrow width approximation i.e. for ${\rm Im}\Pi^R_{\rm vac}\rightarrow 0$,
Eq.~(\ref{gm_Q_fold}) is merged to (\ref{gm_Q}).
The $T$ dependency of 
$\Gamma_Q$ and its corresponding $\tau$ after
folding are shown by solid line in the upper and lower panel of 
Fig.~(\ref{gm_T}) respectively. Due to folding, $\Gamma_Q$
at low $T$ domain (where $m_\pi<2M_Q$) has acquired some non-zero
values from its vanishing contributions and at the same time
corresponding $\tau$ recover from its divergence up to the approximate freeze out
temperature ($T\sim 120-150$ GeV) of the strongly interacting matter.


By using $\Gamma_Q(T,\vk)$ from Eq.~(\ref{gm_Q}) and (\ref{gm_Q_fold})
in the quark component (first term) of Eq.~(\ref{eta_final}), we get
the results of shear viscosity as function of $T$, which are respectively
described by dotted and solid line of Fig.~(\ref{eta_T_comp}). Being
proportional to collisional time, the divergence of $\eta$ is removed after
folding in those temperature region, where $m_\pi<2M_Q$. The contribution
of $\eta$ due to $\Gamma_\pi(T,\vk)$ from Eq.~(\ref{gm_pi}) is shown by
dashed line in Fig.~(\ref{eta_T_comp}). After similar kind of folding
as done in Eq.~(\ref{gm_Q_fold}), an almost negligible ($\sim 10^-5$ GeV$^3$) 
contribution of $\eta$ for pion component can be obtained which is not
included in final results.
 
In low temperature region, $\eta$ is decreasing with increasing of
$T$ which is analogous to the behavior of liquid (From our daily life
experience, we see that the cooking oil behaves like a less viscous
medium when it is heated).   
Whereas in high temperature domain, $\eta$ become an increasing function of $T$
just like a system of gas. 
The magnitude of $\eta$ in our approach is very close to 
the results of Sasaki and Redlich~\cite{Redlich_NPA} (indicated by triangles) 
but underestimated with respect to the earlier estimation in NJL model 
by Zhuang et al.~\cite{Hufner}. The LQCD calculation of 
$\eta$ ($\eta\sim 0.054-0.47$ GeV$^3$ near $T_c$)
by H. B. Meyer~\cite{Meyer} is higher
than all of these calculations.
From the solid line in the lower panel of Fig.~(\ref{gm_T}),
we see that $\tau$ below the $T\sim 160$ MeV exceeds the typical
value of time period ($\sim 30-50$ fm) during which a strongly interacting matter
survive in the labs of heavy ion collisions.
Therefore the estimation of $\eta$ in low
temperature domain is quite higher than the standard 
calculations of $\eta$ of hadronic matter~\cite{Nicola,Weise,SSS}.
The earlier calculations of NJL model~\cite{Redlich_NPA,Hufner} also displayed 
these discrepancy in the hadronic temperature domain.
 
In summary we have investigated the shear viscosity of
strongly interacting matter in the relaxation time
approximation, where quarks with its dynamical mass
may have some non zero Landau damping
because of its various forward and inverse scattering with pions. 
This Landau damping can be obtained from the thermal
field theoretical calculation of quark self-energy for quark-pion loop.
The temperature dependency of shear viscosity is coming from the thermal distribution
functions, the temperature dependence of Landau damping as well as the constituent 
quark mass, supplied by the temperature
dependent gap equation in the NJL model.
Due to this gap equation, this constituent quark mass drops rapidly towards its 
current mass near $T_c$ to restore the chiral
symmetry. A non-trivial influence of all these temperature dependency 
on $\eta(T)$ is displayed in our results.

{\bf Acknowledgment:} S. G. thanks to Saurav Sarkar, Tamal K. Mukherjee, Soumitra Maity, 
Ramaprasad Adak, Kinkar Saha, Sudipa Upadhaya for some pieces of discussions 
which have some direct and indirect influence on our present work.

\end{document}